\documentclass[12pt,a4paper]{article}

\usepackage{amsfonts}

\usepackage{epsfig}
\usepackage{graphics}

\newcommand{\tpretitle}[1]{}
\newcommand{\arttitle}[1]{}

\newcommand{\tnote}[1]{}
\newcommand{\tcomment}[1]{}
\newcommand{\tnotpre}[1]{#1}
\newcommand{\tpre}[1]{}
\newcommand{\tprenote}[1]{}

\newcommand{\href}[2]{#2}
\newcommand{\eprint}[1]{{\tt #1}}

\newcommand{\tsedevelop}[1]{{}}

\renewcommand{\tnotpre}[1]{}
\renewcommand{\tpre}[1]{#1}
\renewcommand{\tprenote}[1]{\footnote{#1}}
\renewcommand{\href}[2]{{#2}{}}
\renewcommand{\eprint}[1]{\href{http://xxx.soton.ac.uk/abs/#1}{{\tt #1}}}

\renewcommand{\tpretitle}[1]{{\em #1}}

\renewcommand{\arttitle}[1]{{\em #1},}

\begin{document}

pre{\begin{flushright} {\tt Imperial/TP/99-0/038} \\
\eprint{hep-ph/0008307} \\
10th August 2000 \\
\tsedevelop{ (LaTeX-ed on \today ) }
\end{flushright}
\vspace*{1cm}}

\begin{center}
{\Large\bf   Scalar fields at finite densities: A $\delta$ expansion approach }
\\ {\tpre{ \vspace*{1cm} } }
{\large
D.Winder\footnote{email: \href{mailto:d.winder@ic.ac.uk} {\tt d.winder@ic.ac.uk}
    \tpre{Tel: [+44]-20-7594-7839,
    Fax: [+44]-20-7594-7844.}}
\\[1cm]
}
\href{http://euclid.tp.ph.ic.ac.uk/}
Theoretical Physics, Blackett Laboratory, Imperial College, \\
Prince Consort Road, London, SW7 2BZ,  U.K.
\end{center}

\vspace*{1cm}

\abstract{
We use an optimized hopping parameter expansion (linear $\delta$-expansion) for the free energy 
to study the phase transitions at finite temperature and 
finite charge density in a global U(1) scalar Higgs sector in the continuum and on the lattice 
at large lattice couplings. We are able to plot out phase diagrams in lattice parameter space 
and find that the standard second-order phase transition with temperature at zero chemical 
potential becomes first order as the chemical potential 
increases.}

\section{Motivation}

In this talk we will tackle phase transitions in the $U(1)$ or $O(2)$ model at finite 
temperature and chemical potential. The work sketched out here can be found 
in the papers \cite{JP}, \cite{EJW} and builds on that set out in \cite{EIM} for the 
case of zero temperature and zero chemical potential. 

We are considering the statistical partition function
\begin{eqnarray}\label{partition}
Z = \mbox{\rm Tr} \left( e^{-\beta H - \beta \mu Q} \right) = \mbox{\rm Tr} \left( e^{-\beta H_{\rm eff}} \right) \nonumber\\
Z =  \int_{\Phi(0)=\Phi(\beta)} {\cal D} \Phi 
	\exp\left\{- \int_{0}^{\beta} d t_4 d^3 \vec{x} {\cal L}_{\rm eff}(\mu)\right\}
\end{eqnarray}
where we have now expressed the partition in a Euclidean path integral representation. For a 
global $U(1)$ scalar field theory the Lagrangian takes the form
\begin{equation}\label{L_eff}
{\cal L}_{\rm eff} = (\nabla_\mu \Phi)^* (\nabla^\mu \Phi)
+ ( m_0^2 - \mu_0^2) \Phi^* \Phi + \frac{\lambda_0}{4} (\Phi^* \Phi)^2
- \mu_0 ( \Phi^* \nabla_4 \Phi - \Phi \nabla_4 \Phi^* )
\end{equation}
To tackle phase transitions in this model we need some non-perturbative techniques.
The obvious first choice is some type of Monte Carlo technique. Unfortunately 
in a statistical integration technique $0 < e^{-S} < 1$ is used as a statistical weight
for each possible field configuration but $S_{\rm eff}$ is {\it complex} and so cannot
be so employed.

We therefore turn to analytic non-perturbative techniques. Both large $N$ approximations and
Hartree Fock resumations are hard to extend beyond leading order; instead we shall use 
a {\it linear delta expansion} (LDE) approach. Firstly we will consider an LDE optimization 
of a $\lambda$ expansion in the continuum and then go on to consider an LDE optimization of
a $\kappa$ expansion on the lattice.

\section{The Linear Delta expansion}

This is an analytic procedure for optimizing a given expansion to give non-perturbative
results. The procedure is expressed in the following steps

(1) Use an {\it interpolating} action:
\begin{equation}\label{S_de}
S \longrightarrow S_{\delta} = S_0(\vec{v}) + \delta ( S - S_0(\vec{v}))
\end{equation}

(2) Expand in powers of $\delta$ and then truncate at $\delta^n$. Setting $\delta=1$ leaves
a residual unphysical dependence on $\vec{v}$. 

(3) Choose values for $\vec{v}$ {\it order by order} in the $\delta$ expansion.
The most popular way of choosing the $\vec{v}$ values is by appling the principle of 
minimal sensitivity (PMS) to some observable.
\begin{eqnarray}\label{PMS_cond}
\frac{\partial \left\langle \mathcal{O} \right\rangle_n}{\partial v^i} = 0
\Longrightarrow \vec{v} = \vec{v}_n \nonumber\\
\left\langle \mathcal{O} \right\rangle_{\rm estimate} = \left\langle \mathcal{O} \right\rangle_n (\vec{v}_n)
\end{eqnarray}
One can think of this as a parametized resummation scheme. If $S_0$ has the same form as $S$
we have a finite, parametized rescaling of physical parameters, that is, an order
by order optimized renormalization scheme choice\cite{PMS}. The technique yields 
non-perturbative information from what is, initially, a perturbative (power series) expansion. 

\section{LDE and $\lambda$ perturbation theory}

Following the LDE proceedure, we choose the ${\cal L}_0$ Lagrangian to be
\begin{equation}\label{L_continuum}
{\cal L}_0 = (\nabla_\mu \Phi)^* (\nabla^\mu \Phi)
+ ( \Omega^2 - \mu_0^2) \Phi^* \Phi 
-\mu_0 ( \Phi^* \nabla_4 \Phi - \Phi \nabla_4 \Phi^* ) 
\end{equation}
This gives an LDE interpolating lagrangian of the form
\begin{eqnarray}\label{L_de}
{\cal L}_{\delta} = (\nabla_\mu \Phi)^* (\nabla^\mu \Phi)
+ ( \Omega^2 - \mu_0^2) \Phi^* \Phi 
- \mu_0 ( \Phi^* \nabla_4 \Phi - \Phi \nabla_4 \Phi^* ) \nonumber\\
+ \delta \left[- ( \Omega^2 - m_0^2) \Phi^* \Phi + \frac{\lambda_0}{4} (\Phi^* \Phi)^2 \right]
\end{eqnarray}
In terms of the partition we have
\begin{equation}\label{partition_LDE}
Z = \exp\left\{ \delta \left[ -(\Omega^2 - m_0^2) \int d t \frac{ \delta^2}{\delta j \delta j^*} 
+ \frac{\lambda_0}{4} \int d t \frac{ \delta^4}{\delta j^2 \delta j^{* 2}} \right] \right\} Z_0[j]
\end{equation}
Performing the expansion for the self energy one sees that the expansion is optimized
by a `mass insertion' term 
\setlength{\unitlength}{0.1cm}
$\begin{picture}(10,1)
\put(1,1){\circle*{0.5}}
\put(1,1){\line(1,0){8}}
\put(5,1){\circle*{1}}
\put(9,1){\circle*{0.5}}
\end{picture}$. The self energy to $O(\delta^2)$ is 
\setlength{\unitlength}{0.09cm}
\begin{equation}\label{continuum_expansion}
\begin{picture}(10,5)
\put(1,1){\circle*{0.5}}
\put(1,1){\line(1,0){8}}
\put(5,1){\circle*{4}}
\put(9,1){\circle*{0.5}}
\end{picture} 
=
\begin{picture}(10,5)
\put(1,1){\circle*{0.5}}
\put(1,1){\line(1,0){8}}
\put(9,1){\circle*{0.5}}
\end{picture} 
+
\begin{picture}(10,5)
\put(1,1){\circle*{0.5}}
\put(1,1){\line(1,0){8}}
\put(5,2.6){\circle{3}}
\put(9,1){\circle*{0.5}}
\end{picture} 
+ 
\begin{picture}(10,5)
\put(1,1){\circle*{0.5}}
\put(1,1){\line(1,0){8}}
\put(5,1){\circle*{1}}
\put(9,1){\circle*{0.5}}
\end{picture} 
+
\begin{picture}(10,5)
\put(1,1){\circle*{0.5}}
\put(1,1){\line(1,0){8}}
\put(5,2.6){\circle{3}}
\put(5,5.7){\circle{3}}
\put(9,1){\circle*{0.5}}
\end{picture} 
+
\begin{picture}(10,5)
\put(1,1){\circle*{0.5}}
\put(1,1){\line(1,0){8}}
\put(3,2.6){\circle{3}}
\put(7,2.6){\circle{3}}
\put(9,1){\circle*{0.5}}
\end{picture} 
+
\begin{picture}(10,5)
\put(1,1){\circle*{0.5}}
\put(1,1){\line(1,0){8}}
\put(5,1){\circle{4}}
\put(9,1){\circle*{0.5}}
\end{picture}
+
\begin{picture}(10,5)
\put(1,1){\circle*{0.5}}
\put(1,1){\line(1,0){8}}
\put(5,2.6){\circle{3}}
\put(5,4){\circle*{1}}
\put(9,1){\circle*{0.5}}
\end{picture} 
+
\begin{picture}(10,5)
\put(1,1){\circle*{0.5}}
\put(1,1){\line(1,0){8}}
\put(3,1){\circle*{1}}
\put(7,1){\circle*{1}}
\put(9,1){\circle*{0.5}}
\end{picture} 
\end{equation}
Renormalization is done using counterterms and this introduces a renormalization scale $M^2$.
All the resulting integrals are evaluated using a high $T$ expansion. The thermal mass $m_T^2$
is calculated, and one then fixes the variational parameter 
using $\frac{\partial m_T^2}{\partial \eta} = 0 $, where $\eta^2 = \Omega^2 - m_0^2$.

\begin{figure}[hbtp]
\begin{center}
\scalebox{0.71}{\includegraphics*[3cm,19.5cm][19.5cm,28cm]{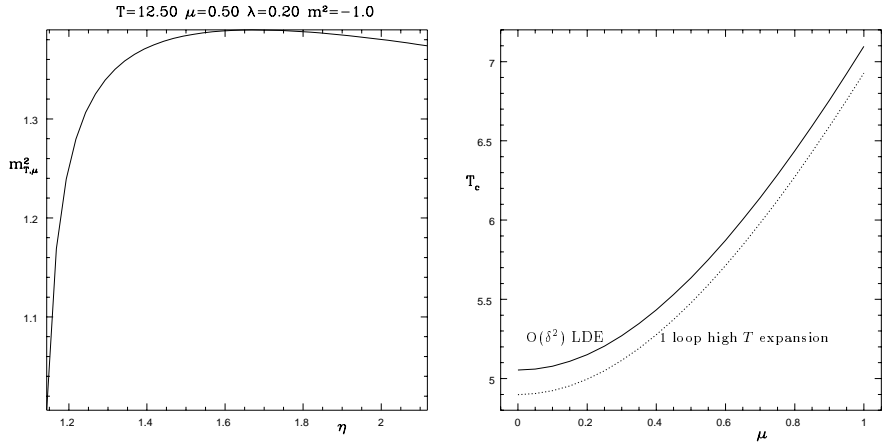}}
\caption{(a) Minimization curve for $m_T^2$. (b) Phase transition curve in $\{T,\mu\}$ space.}
\label{continue}
\end{center}
\end{figure}

An typical minimization plot for $m_T^2$ is seen in Figure \ref{continue}, along with the 
resulting phase transition curves in $\{ T,\mu\}$ space as compared to the $1$-loop high 
temperature approximation.

\section{LDE and $\kappa$ perturbation theory}

\begin{figure}[htbp]
\begin{center}
\scalebox{0.67}{\includegraphics*[2.8cm,17.7cm][19.5cm,27cm]{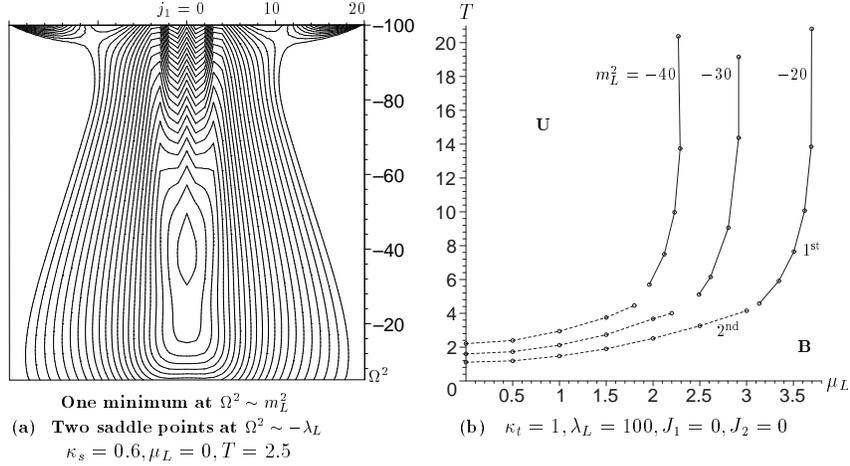}}
\caption{(a) Example minimization contour plot in unbroken regime. 
	(b) Phase transition curves in $\{T,\mu\}$ space. Note that the phase transition 
	becomes first order for large $\mu_L$.}
\label{discrete}
\end{center}
\end{figure}

The discretized Lagrangian on an {\it asymmetric} lattice 
(different spatial and temporal lattice spacings; $a_s, a_t$) is expressed in
terms of dimensionless physical parameters and has the form
\begin{equation}\label{L_discrete}
L_n =
-\sum_{i}
\begin{picture}(10,1)
\put(1,1){\circle*{0.5}}
\put(1,1){\line(1,0){8}}
\put(9,1){\circle*{0.5}}
\end{picture}
-
\begin{picture}(4,8)
\thicklines
\put(2,-3){\circle*{0.5}}
\put(2,-3){\vector(0,1){8}}
\put(2,5){\circle*{0.5}}
\end{picture}
+ (m_L^2 - \mu_L^2) \Phi_n^* \Phi_n
+ \lambda_L ( \Phi_n^* \Phi_n )^2 - J_L^{*} \Phi_n - J_L \Phi_n^* 
\end{equation}
where $n \equiv (n_4, \bf{n})$ and the spatial and temporal links are defined by
\begin{eqnarray}\label{s_and_t_links}
\begin{picture}(10,1)
\put(1,1){\circle*{0.5}}
\put(1,1){\line(1,0){8}}
\put(9,1){\circle*{0.5}}
\end{picture} 
&:=&
\kappa_s \left[ \Phi_{n_4,\bf{n}}^* \Phi_{n_4,\bf{n}+\bf{e}_i} +
\Phi_{n_4,\bf{n}} \Phi_{n_4,\bf{n}+\bf{e}_i}^* \right]
\nonumber\\
\begin{picture}(4,8)
\thicklines
\put(2,-3){\circle*{0.5}}
\put(2,-3){\vector(0,1){8}}
\put(2,5){\circle*{0.5}}
\end{picture}
&:=&
\left[ \kappa_{t} \left(1 + \mu_L \right) \Phi_{n_4,\bf{n}}^* \Phi_{n_4+1,\bf{n}}
+ \kappa_{t} \left (1 - \mu_L \right) \Phi_{n_4,\bf{n}} \Phi_{n_4+1,\bf{n}}^* \right] 
\end{eqnarray}
The $L_0$ Lagrangian is
\begin{equation}\label{L_d0}
L_{0 n} =
(\Omega^2 - \mu_L^2) \Phi_n^* \Phi_n 
+ \lambda_L ( \Phi_n^* \Phi_n )^2 - j^{*} \Phi_n - j \Phi_n^* 
\end{equation}
Thus the LDE Lagrangian has the form
\begin{eqnarray}\label{L_de_l}
L_{\delta} &&=
(\Omega^2 - \mu_L^2) \Phi_n^* \Phi_n 
+ \lambda_L ( \Phi_n^* \Phi_n )^2 - j^{*} \Phi_n - j \Phi_n^* 
+ \delta \left[
-\sum_{i}
\begin{picture}(10,1)
\put(1,1){\circle*{0.5}}
\put(1,1){\line(1,0){8}}
\put(9,1){\circle*{0.5}}
\end{picture}
-
\begin{picture}(4,8)
\thicklines
\put(2,-3){\circle*{0.5}}
\put(2,-3){\vector(0,1){8}}
\put(2,5){\circle*{0.5}}
\end{picture}
\right] \nonumber\\
&&+\delta \left[
- (\Omega^2 - m_L^2) \Phi_n^* \Phi_n 
- (J_L^*- j^{*}) \Phi_n - (J_L - j) \Phi_n^* \right]
\end{eqnarray}
The free energy expansion can be expressed in terms of {\it cumulant} averages with respect
to the {\it ultra local} Lagrangian, $L_U$. This is the same as the $L_\delta$ Lagrangian 
but without the spatial and temporal link terms (\ref{s_and_t_links}). The free energy is
\setlength{\unitlength}{0.1cm}
\begin{equation}\label{F}
F = - \frac{1}{N} \ln Z_U
- \frac{1}{N} \sum_{j=1}^{\infty} \frac{\delta^j}{j!} \left\langle
\left(
\sum_{n}
\begin{picture}(4,8)
\thicklines
\put(2,-3){\circle*{0.5}}
\put(2,-3){\vector(0,1){8}}
\put(2,5){\circle*{0.5}}
\end{picture}
+\sum_{n,i}
\begin{picture}(10,1)
\put(1,1){\circle*{0.5}}
\put(1,1){\line(1,0){8}}
\put(9,1){\circle*{0.5}}
\end{picture} \right)^j
\right\rangle_C
\end{equation}
and therefore has a diagrammatic representation in terms of connected spatial and temporal
links. The variational parameters, $j$ and $\Omega^2$, are fixed using the PMS conditions
\begin{equation}\label{PMS_lattice}
\frac{\partial F}{\partial j} = \frac{\partial F}{\partial \Omega^2} = 0
\end{equation}
and the phase transition is tracked 
using $\left. \left( \partial F / \partial J_1 \right) \right|_{J_1 = 0} 
= - \left\langle \Phi_1 \right\rangle$.
A typical minimization contour plot is seen in Figure \ref{discrete}, along with the 
resulting phase transition curves in $\{ T,\mu_L\}$ space. Note that the phase transition 
becomes first order for sufficently high values of $\mu_L$.

\section{Conclusions}\label{conclusions}

An LDE optimization of the standard $\lambda$ and hopping parameter expansions has allowed 
access to some of the truly non-perturbative physics of the global scalar $U(1)$ model 
at finite $T$ and $\mu_L$. On the lattice one finds a {\it first order} phase transition 
at sufficiently large $\mu_L$. The approach can be extended to plotting out phase diagrams 
at finite density for a more complex theory, e.g. gauge theory.

\section*{Acknowledgements}

The work discussed in this talk was completed in collaboration with H.~F. Jones, T.~S. Evans 
and P. Parkin.



\begin{thebibliography}{99}

\bibitem{JP}
H.F.~Jones and P.~Parkin,
[hep-th/0005069].

\bibitem{EJW}
T.S.~Evans, H.F.~Jones and D.~Winder,
[hep-th/0008307].

\bibitem{EIM}
T.S.~Evans, M.~Ivin and M.~M\"obius,
Nucl.~Phys.{\bf B577} (2000) 325.

\bibitem{PMS}
P.M.~Stevenson,
Phys.~Rev. D {\bf 23} (1981) 2916.

\end{thebibliography}
\end{document}